\documentclass[letterpaper]{article} 
\usepackage[submission]{aaai25}  
\usepackage{times}  
\usepackage{helvet}  
\usepackage{courier}  
\usepackage[hyphens]{url}  
\usepackage{graphicx} 
\usepackage{natbib}  
\usepackage{caption} 
\usepackage{algorithm}
\usepackage{algorithmic}

\usepackage{multirow}
\usepackage{booktabs} 
\usepackage{makecell}
\usepackage{amssymb}

\usepackage{amsmath}

\usepackage{xcolor}

\usepackage{newfloat}
\usepackage{listings}
\DeclareCaptionStyle{ruled}{labelfont=normalfont,labelsep=colon,strut=off} 
\floatstyle{ruled}
\newfloat{listing}{tb}{lst}{}
\floatname{listing}{Listing}
\title{Diff-Cleanse: Identifying and Mitigating Backdoor Attacks in Diffusion Models}
\author{
    Hao Jiang\textsuperscript{\rm 1},
    Jin Xiao\textsuperscript{\rm 1},
    Xiaoguang Hu\textsuperscript{\rm 1},
    Tianyou Chen\textsuperscript{\rm 1},
    Jiajia Zhao\textsuperscript{\rm 2}
}
\affiliations{
    \textsuperscript{\rm 1}Beihang University\\
    \textsuperscript{\rm 1}National Key Laboratory of Complex System Control and Intelligent Agent Cooperation
}

\usepackage{bibentry}

\begin{document}

\maketitle

\begin{abstract}
Diffusion models (DMs) are regarded as one of the most advanced generative models today, yet recent studies suggest that they are vulnerable to backdoor attacks, which establish hidden associations between particular input patterns and model behaviors, compromising model integrity by causing undesirable actions with manipulated inputs. This vulnerability poses substantial risks, including reputational damage to model owners and the dissemination of harmful content. To mitigate the threat of backdoor attacks, there have been some investigations on backdoor detection and model repair. However, previous work fails to reliably purify the models backdoored by state-of-the-art attack methods, rendering the field much underexplored. To bridge this gap, we introduce Diff-Cleanse, a novel two-stage backdoor defense framework specifically designed for DMs. The first stage employs a novel trigger inversion technique to reconstruct the trigger and detect the backdoor, and the second stage utilizes a structural pruning method to eliminate the backdoor. We evaluate our framework on hundreds of DMs that are attacked by three existing backdoor attack methods with a wide range of hyperparameter settings. Extensive experiments demonstrate that Diff-Cleanse achieves nearly 100\% detection accuracy and effectively mitigates backdoor impacts, preserving the model's benign performance with minimal compromise.
\end{abstract}

%

\section{Introduction}
Diffusion models (DMs) have emerged as one of the most successful generative models, achieving superior performance \cite{dhariwal2021diffusion} and finding widespread applications in image \cite{nichol2021glide}, audio \cite{kong2020diffwave}, video \cite{mei2023vidm,yang2023video}, text \cite{li2022diffusion}, and text-to-speech \cite{jeong2021diff,huang2022fastdiff}
generation. Moreover, diffusion models are increasingly used in safety-critical tasks, such as autonomous driving \cite{goelimproving} and medical imaging \cite{wolleb2022diffusion,moghadam2023morphology}.

However, recent studies show that DMs are vulnerable to backdoor attacks \cite{chou2023backdoor,chen2023trojdiff,chou2023villandiffusion}, where attackers poison a portion of the training dataset, causing the model to induce model behavior triggered by specific input patterns. Given the common practice of using powerful pre-trained models from online sources or outsourcing the training process to third parties, such attacks could have disastrous consequences \cite{chou2023backdoor}.
For instance, a company using a compromised DM for image generation service might inadvertently produce harmful content, damaging the company's reputation and leading to legal repercussions \cite{an2024elijah}. Furthermore, DMs are often used to generate synthetic data for training downstream models in scenarios with limited data or privacy concerns \cite{liu2019ppgan,goelimproving}. Attackers could implant biases within these models and affect downstream applications.

Despite these challenges, there is a significant lack of backdoor defenses for DMs. Most existing backdoor defense methods are designed for discriminative models like classifiers and are not directly applicable to DMs due to different prediction objectives (e.g., class probability for classifiers, mean or noise estimation for DMs) and inference processes. Furthermore, current defense approaches for DMs are limited, either focusing solely on input-level detection \cite{guan2024ufid,sui2024disdet} or failing to effectively identify backdoors in our experiments \cite{an2024elijah}.

To address this gap, we investigate three state-of-the-art backdoor attacks on DMs in the context of unconditional noise-to-image generation. Despite variations in trigger and target patterns, attackers typically seek models with high utility and high specificity \cite{chou2023backdoor}. High utility refers to a model’s ability to generate high-quality and diverse images from clean inputs, while high specificity means the model produces images matching the target distribution when presented with inputs containing trigger patterns. To mitigate the backdoor, a natural approach involves synthesizing an effective trigger and removing the channels activated by it. Our experiments show that pruning a small number of channels (typically $1\% \sim 2\%$) is sufficient to remove backdoors. Based on these insights, we propose a two-stage backdoor defense framework for DMs. First, we introduce a novel trigger inversion method that formulates trigger inversion as an optimization problem, finding a trigger that reduces the diversity of generated images. We develop a metric, Max Similarity Cluster Ratio (MSCR), to measure the diversity of images generated with the inverted trigger and use this metric as an indicator of backdoors. Then, to eliminate the backdoor, we propose a pruning-based method to remove backdoor-related channels and design a loss function to enhance fine-tuning effectiveness with a limited number of clean samples. In summary, the main contributions of this paper are as follows. 

\begin{itemize}

\item We present Diff-Cleanse, a two-stage backdoor defense framework for diffusion models (DMs), which can effectively detect and mitigate backdoors created by state-of-the-art attack methods. 

\item We introduce a novel Black Box Trigger Inversion method (BBTI) based on the pursuit of high specificity in existing backdoor attacks. This method includes a new metric, the Max Similarity Cluster Ratio (MSCR), to assess image diversity and detect backdoors.

\item We devise a simple yet effective pruning-based backdoor removal algorithm, incorporating a novel loss function, to eliminate backdoors. To the best of our knowledge, this is the first study to explore the application of pruning methods in the field of diffusion model backdoor defense.

\item Extensive experiments on 177 clean and 196 backdoored models demonstrate that Diff-Cleanse achieves close to 100\% detection accuracy, reduces backdoor effects to nearly zero, and largely maintains the model's utility.

\end{itemize}

\section{Related Works}
\label{sec:relatedworks}

\subsection{Diffusion Models}
\label{subsec:diffusoinmodels}

Diffusion models (DMs) are a class of generative model that learns the reversed diffusion process which is derived from a tractable forward corruption process \cite{chou2023villandiffusion}. Utilizing DMs generally involves three key procedures. (1) \textbf{Diffusion process.} Define a diffusion process $q\left(\boldsymbol{x}_{t} \mid \boldsymbol{x}_{t-1}\right)$ that diffuses the data distribution $q(\boldsymbol{x})$ into a certain distribution $\tau(\boldsymbol{x})$ over $T$ timesteps. (2) \textbf{Training.} Train parameters $\theta$ to ensure that the generative process aligns with the reversed diffusion process, i.e., $p_\theta\left(\boldsymbol{x}_{t-1} \mid \boldsymbol{x}_t\right)=\mathcal{N}\left(\boldsymbol{x}_{t-1};\mu_\theta\left(\boldsymbol{x}_t,t\right),\beta_\theta\left(\boldsymbol{x}_t,t\right)\right)=q\left(\boldsymbol{x}_{t-1} \mid \boldsymbol{x}_t\right)$. (3) \textbf{Sampling.} Use the trained generative process $p_{\theta}\left(\boldsymbol{x}_{t-1} \mid \boldsymbol{x}_t\right)$ to sample from $t=T_{\max}$ down to $T_{\min}$. We provide a more detailed introduction in appendix A.

\begin{figure}[htb]
\centering
\includegraphics[width=0.8\linewidth]{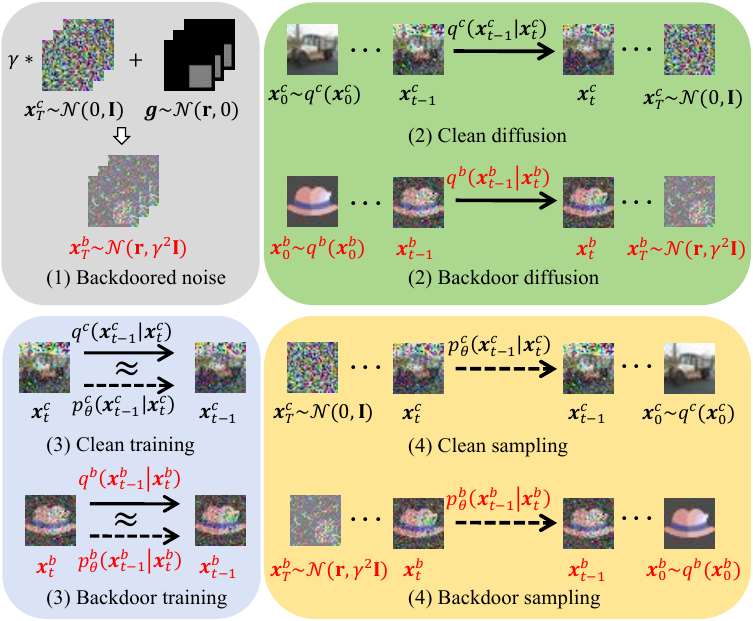}
\caption{An illustration of backdoor attacks on unconditional noise-to-image diffusion models, with “Gray Box” and “Hat” as examples of the trigger and target respectively. (1) The backdoored noise $\boldsymbol{x}_{T}^b$ is the weighted sum of the trigger $\boldsymbol{g}$ and the clean noise $\boldsymbol{x}_{T}^c$. (2) The clean diffusion process transforms the dataset distribution $q^c\left(\boldsymbol{x}_0^c\right)$ into the standard normal distribution $\mathcal{N}(0, \mathbf{I})$, while the backdoor diffusion process transforms the target distribution $q^b\left(\boldsymbol{x}_0^b\right)$ into a shifted and rescaled normal distribution $\mathcal{N}\left(\mathbf{r}, \gamma^2 \mathbf{I}\right)$. (3) Druing training, parameters $\theta$ learn both the clean reverse process $p_\theta^c\left(\boldsymbol{x}_{t-1}^c \mid \boldsymbol{x}_t^c\right)$ and the backdoor reverse process $p_\theta^b\left(\boldsymbol{x}_{t-1}^b \mid \boldsymbol{x}_t^b\right)$. (4) After training, the backdoored model cna sample normal images $\boldsymbol{x}_0^{c}\sim q^c\left(x_0^c\right)$ from the clean noise, and target images $\boldsymbol{x}_0^{b}$ from the backdoored noise.}
\label{fig:backdoorattack}
\end{figure}

\subsection{Backdoor Attacks on Diffusion Models}
\label{subsec:backdoorattacks}
Injecting backdoors to discriminative models (e.g., classifiers) involves poisoning a small portion of the training data by stamping the predefined trigger and relabeling them as target classes \cite{chen2017targeted,gu2019badnets,nguyen2020input,xue2020one}. After training, the model would produce the intended target label in the presence of the trigger. Backdoor attacks on diffusion models are more complex, as the sampling process depends on both the input and the transitional distribution. We categorize backdoor attacks on DMs into those targeting conditional and unconditional DMs. Recent studies indicates that conditional DMs can be backdoored through natural language prompts \cite{zhai2023text,struppek2023rickrolling,huang2023zero} or by poisoning the training set \cite{pan2023trojan}. These attacks can be countered by checking the text encoder or the distribution of generated images \cite{sui2024disdet}. Therefore, this paper focuses on pixel-level backdoor attacks designed for unconditional DMs, which have better stealth and pose a broader threat.

To the best of our knowledge, there are three existing backdoor attacks for unconditional DMs, BadDiff \cite{chou2023backdoor}, TrojDiff \cite{chen2023trojdiff} and VillanDiff \cite{chou2023villandiffusion}. Their workflow is summarized in Fig. \ref{fig:backdoorattack}. Attackers first define a forward backdoor diffusion process $\boldsymbol{x}_0^b \rightarrow \boldsymbol{x}_T^b$, where $\boldsymbol{x}_0^b$ is the target image and $\boldsymbol{x}_T^b$ is the backdoored noise. Without loss of generality, we assume $\boldsymbol{x}_T^b=\boldsymbol{g}+\gamma\cdot \boldsymbol{x}_T^c$, where $\boldsymbol{g}\sim\mathcal{N}(\mathbf{r}, 0)$ is the trigger, $\gamma$ is the the weight coefficient and $\boldsymbol{x}_T^c\sim\mathcal{N}(0,\mathbf{I})$ is the clean input. Attackers then derive the corresponding backdoor reverse process and the loss of backdoor training. During training, clean and backdoor losses are calculated separately and the parameter $\theta$ learns the clean transitional distribution $q^c(\boldsymbol{x}_{t-1}^c\mid \boldsymbol{x}^c_t)$ and the backdoor transitional distribution $q^{b}(\boldsymbol{x}^{b}_{t-1}\mid \boldsymbol{x}^{b}_t)$. During sampling, the backdoored model generates clean samples $\boldsymbol{x}_0^c\sim q^c(\boldsymbol{x}_0^c)$ from clean input, and target images $\boldsymbol{x}_0^b\sim q^b(\boldsymbol{x}_0^b)$ from backdoored noise. 

Among these attacks, TrojDiff introduces three distinct attacks, each with a different type of target distribution:
\begin{itemize}
\item \textbf{D2I Attack.} $q^b(\boldsymbol{x}_0^b)=x_{\text{target}}$, where $x_{\text{target}}$ is a predefined target image, such as “Mickey Mouse”.

\item \textbf{Din Attack.}  $q^b(\boldsymbol{x}_0^b)=q(x \mid \hat{y})$, where $\hat{y}$ is a target class in the class set of $q^c(\boldsymbol{x}_0^c)$. For instance, use CIFAR-10 as the clean training dataset and “horse” as the target class.

\item \textbf{Dout Attack.}  $q^b(\boldsymbol{x}_0^b)=q(x \mid \hat{y})$, where $\hat{y}$ is a target class outside the class set of $q^c(\boldsymbol{x}_0^c)$. For example, use CIFAR-10 as the clean training dataset and the number “7” from MNIST as the target class.
\end{itemize}

BadDiff and VillanDiff only implement the D2I attack. BadDiff is limited to the ancestral sampler DDPM, while VillanDiff proposes a general framework applicable to advanced samplers and more types of diffusion models, such as latent diffusion models \cite{rombach2022high} and score-based diffusion models \cite{song2020score}. TrojDiff encompasses all three types of attack and designs backdoor sampling functions for DDPM and DDIM sampler.

\subsection{Existing Backdoor Defenses}
\label{subsec:existingdefenses}
Due to different prediction objectives and inference processes, defense methods for other tasks cannot be directly applied to DMs \cite{an2024elijah}. Recent efforts to mitigate backdoor impacts in DMs can be categorized into input-level and model-level approaches. Input-level defenses focus on detecting whether an input sample is poisoned. For instance, UFID \cite{guan2024ufid} introduces a framework that identifies backdoored inputs by analyzing the sensitivity of input-output pairs to perturbations. DisDet \cite{sui2024disdet} distinguishes poisoned noise from clean Gaussian noise through data distribution analysis. Model-level defenses aim to detect and mitigate backdoors within the models. Elijah \cite{an2024elijah} proposes a trigger inversion method based on distribution shift preservation and removes the backdoor by aligning predictions for clean and backdoored noise. TERD \cite{moterd} designs a novel trigger inversion algorithm and develops input-level and model-level detection methods based on high-quality reversed triggers. Notably, only Elijah explores backdoor removal in DMs, but it is limited to defending against the D2I attack.

\begin{figure*}[tb]  
\centering  
\includegraphics[width=0.66\textwidth]{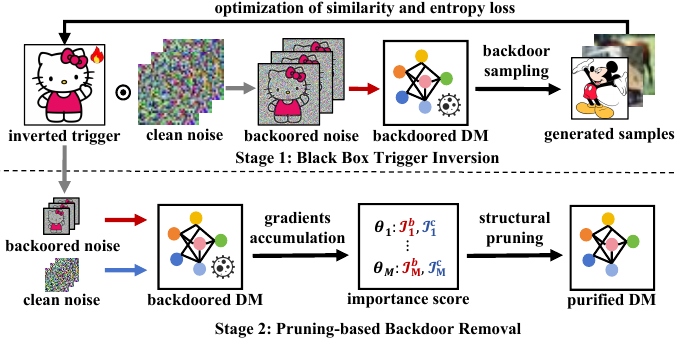}  
\caption{Overview of our two-stage backdoor defense framework  Diff-Cleanse with “Mickey” and “Hello Kitty” as examples of the trigger and the target respectively. Stage 1 reconstructs the trigger and detects the backdoor. Stage 2 removes the backdoor.}
\label{fig:frameworkoverview}  
\end{figure*}

\section{Methodology}

This section first introduces the threat model that defines the scope and property of the backdoor attacks considered in this study. We then outline our proposed two-stage backdoor defense framework, as illustrated in Fig. \ref{fig:frameworkoverview}. Given a risky diffusion model (DM), we first perform a trigger inversion algorithm to reconstruct the trigger and detect any backdoor. If detected, we apply a structural pruning-based algorithm to remove it. In the following sections, $\mathbf{r}\in \mathbb{R}^{H \times W \times 3}$ denotes the specific value of trigger $\boldsymbol{g}$ and $\boldsymbol{X}_t\in \mathbb{R}^{N\times H \times W \times 3}$ denotes $N$ sampled images at timestep $t$. $H$ and $W$ represent the image height and width respectively.

\subsection{Threat Model}
We adopt a consistent threat model with existing literature \cite{wang2019neural,an2024elijah} in the context of unconditional noise-to-image diffusion models, which describes the goals and capacities of two parties, attacker and defender. 

\paragraph{Attacker.} The attacker's goal is to develop a backdoored DM with two properties \cite{chou2023backdoor}: (1) High utility. The model performs well in clean sampling.
(2) High specificity. Activating the backdoor causes the model to generate target images. In tersm of capability, we assume strong attackers who can (1) design the backdoor diffusion and sampling process, (2) modify all aspects of training, such as loss function, training dataset and hyperparameters.

\paragraph{Defender.} The defender's goals are to (1) detect potential backdoors, and (2) remove them while preserving the models' benign utility. The defender is assumed to have no prior knowledge of the attack settings but has white-box access to the model and a limited amount of clean data from the original training dataset. Notably, our framework works without any real clean data. The trigger inversion and backdoor detection method require no image samples. The backdoor removal method, though requiring clean data, performs well with synthetic data generated by the compromised DM, achieving competitive results compared to using real data. 

\subsection{Overview of Diff-Cleanse}

\textbf{Key intuition} The intuition behind our technique stems from the basic characteristics of backdoor attacks: clean DMs convert Gaussian noise into the clean dataset distribution via clean sampling, while backdoored DMs can additionally convert backdoored noise into the target distribution via backdoor sampling. Recall that in attacks addressed in this paper, the trigger causes backdoored noise to deviate significantly from clean noise, reducing sampling diversity and producing identical or similar outputs within the same class. Intuitively, we hypothesize: (1) trigger patterns that reduce sampling diversity can aid in backdoor detection, and (2) channels activated by the inverted trigger will also respond to the attacker-specified trigger. These hypotheses lead to the following observations.


\textbf{Observation 1} Both clean and backdoor DMs can be triggered to reduce sampling diversity, but the sampling results are visually distinct. Different triggers vary in their effectiveness at reducing diversity.

\begin{figure}
\centering
\includegraphics[width=0.8\linewidth]{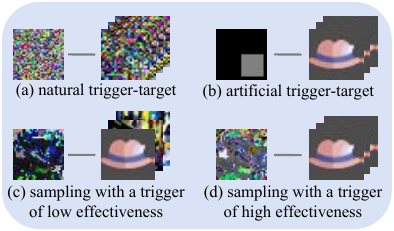}
\caption{Examples of natural (a) and artificial (b) trigger-target pairs. (c) and (d) show the results of sampling using inverted triggers of low and high effectiveness respectively. An effective trigger leads to a larger proportion of the target "Hat" in the generated images.}
\label{fig:naturalartificialtarget}
\end{figure}

Here, we refer to the set of similar samples generated by a trigger as the “target”, and refer to the corresponding pair as the “trigger-target pair”. We categorize trigger-target pairs as either natural or artificial, corresponding to natural and artificial backdoors. Natural triggers and targets emerge during normal training, which are typically images incomprehensible to humans. Artificial triggers and targets are usually human-understandable images. While both clean and backdoored DMs have natural backdoors, only backdoored DMs contain artificial ones. Artificial triggers can be manipulated in aspects like color, shape, and texture. Therefore, we focus on the effectiveness of triggers in activating the backdoor and producing the target. Fig. \ref{fig:naturalartificialtarget} shows examples of both natural and artificial trigger-target pairs, along with backdoor sampling results using triggers of varying effectiveness.


\textbf{Observation 2} Structural pruning method can completely eliminate the backdoor in diffusion models.

We use pruning-based methods to remove backdoors, a common approach in defenses for discriminative models \cite{wang2019neural,wu2021adversarial,zheng2022data}. Existing pruning-based methods primarily target deep layers and using unstructural pruning \cite{liu2018fine,wu2021adversarial}, which minimally affects the model's benign performance. However, we observe that optimal pruning rates vary significantly across attack scenarios when pruning is confined to deep layers, making the approach sensitive to hyperparameters. Additionally, unstructural pruning often fails to completely eliminate the backdoors, as they tend to re-emerge after fine-tuning. In contrast, structural pruning effectively suppresses these phenomena, as shown in Tab. \ref{tab:backdoorremoval} and appendix F.

Based on these observations, we introduce the two-stage backdoor defense framework \textbf{Diff-Cleanse}. The first stage models trigger inversion as an optimization problem to reconstruct effective triggers, which are also used to detect backdoors. If a backdoor is detected, the second stage applies structural pruning to remove backdoor-related channels and restore the model's benign utility using our proposed loss function. We provide the pseudocode in appendix C.

\subsection{Trigger Inversion (Stage 1)}

Recall that the backdoor diffusion maps the target image $x_0^b \sim q^b\left(x_0^b\right)$ to the backdoored noise $\boldsymbol{x}_{T}^{b}\sim \mathcal{N}\left(\mathbf{r}, \gamma^2\mathbf{I}\right)$. Without loss of generality, we consider a unified form of backdoor diffusion as follows.
\begin{equation}
\boldsymbol{x}_t^{b}=\hat{\alpha}_t \boldsymbol{x}_0^{b}+\hat{\beta}_t \boldsymbol{\epsilon}_t+\hat{\rho}_t \mathbf{r}
\label{eq:backdoordiff}
\end{equation}
where $\boldsymbol{\epsilon}_t\sim\mathcal{N}(0, \mathbf{I})$ is Gaussian noise, $\left\{\hat{\alpha}_t\right\}_{t=1}^T$, $\left\{\hat{\rho}_t\right\}_{t=1}^T$ and $\{\hat{\beta}_t\}_{t=1}^T$ are sequences set by the attacker, depending on timestep $t$. $\mathbf{r}$ is the specific value of trigger $\boldsymbol{g}$. Due to space limitations, specific parameter settings for different attacks are provided in appendix B.

Given $t=T$, Eq. \ref{eq:backdoordiff} reveals a linear relationship between the backdoored noise $\boldsymbol{x}^b_T$ and $\mathbf{r}$. Leveraging this, we formulate trigger inversion as an optimization problem with two objectives. First, since attackers seek high specificity in models \cite{chou2023backdoor}, artificial triggers should effectively activate the backdoor, producing attacker-specified target images. Let $\boldsymbol{g}^s\sim \mathcal{N}(\mathbf{r}^s, 0)$ denote the inverted trigger and $\boldsymbol{X}^{s}$ the results of backdoor sampling with $\boldsymbol{g}^s$. Using a pretrained image encoder $\mathcal{E}(\cdot)$, we map $\boldsymbol{X}^{s}$ to embeddings $\boldsymbol{Z}^{s}$, where $\boldsymbol{Z}^{s}=\mathcal{E}(\boldsymbol{X}^{s})$. The encoder retains class discrimination, mapping similar images close in embedding space. To assess the diversity of $\boldsymbol{X}^{s}$, we propose the similarity loss $\ell_{s}$ as in Eq. \ref{eq:ls}.
\begin{equation}
\label{eq:ls}
\ell_{s}(\boldsymbol{g}^{s})=1-\frac{\sum_{i,j} \mathcal{A}(\boldsymbol{Z}^{s}_{i}, \boldsymbol{Z}^{s}_{j})}{|\boldsymbol{Z}^{s}|^2}
\end{equation}

Here, $\left(\boldsymbol{Z}^{s}_i, \boldsymbol{Z}^{s}_j\right)$ are sample pairs from $\boldsymbol{Z}^{s}$. $\mathcal{A}(\cdot)$ is the cosine similarity function and $\left|\cdot\right|$ is the set's cardinality.

\begin{figure}  
\centering
\includegraphics[width=0.6\linewidth]{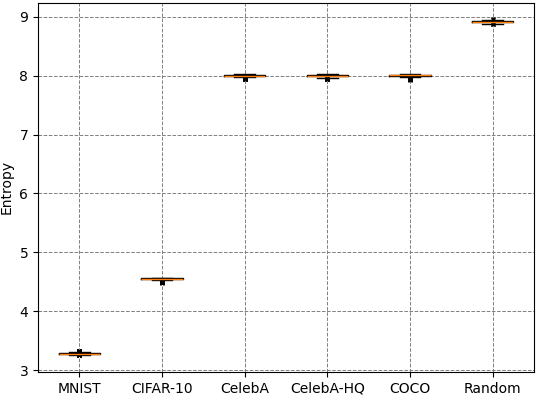}
\caption{The boxplot of entropy values for images in several visual datasets. “Random” refers to 30k images sampled from the Gaussian distribution.}
\label{fig:entropyboxplot}
\end{figure}

The second objective is to identify artificial trigger-target pairs rather than natural trigger-target pairs. Artificial targets, designed by attackers, are assumed to have human-understandable semantics, resulting in entropy within a specific range. In contrast, natural targets typically have either very high (e.g., random noise) or very low (e.g., a pure black image) entropy. As shown in Fig. \ref{fig:entropyboxplot}, different datasets exhibit varying entropy values. Complex textures (e.g., faces in CelebA-HQ) have higher entropy than simple ones (e.g., handwritten digits in MNIST). Random noise has the highest entropy. Thus, we propose the entropy loss $\ell_e$ as follows.
\begin{equation}
\label{eq:le}
\ell_{e}(\boldsymbol{g}^s) = \sum_{i} \left[ \max(\mathcal{H}_i - \kappa_{+}, 0) + \max(\kappa_{-} - \mathcal{H}_i, 0) \right]
\end{equation}
where $\mathcal{H}_i=\mathcal{H}(\boldsymbol{X}^{s}_i)$ is the entropy of the $i$-th sample from $\boldsymbol{X}^s$. The values for $\kappa_{+}$ and $\kappa_{-}$ are not strict and we use 8.5 and 0.1. The final objective combines both losses:
\begin{equation}
\min_{\boldsymbol{g}^s} \ell\left(\boldsymbol{g}^s\right)=\ell_{s}\left(\boldsymbol{g}^s\right)+\lambda \ell_{e}\left(\boldsymbol{g}^s\right)
\label{eq:l(g)}
\end{equation}
where $\lambda$ is the weight of $\ell_{e}$. In experiments, we adjust $\lambda$ dynamically so that the two loss terms are of similar magnitude. Considering $\hat{\rho}_T=1$, we use Eq. \ref{eq:backdoordiff} to calculate the derivative of $\boldsymbol{x}_0^s$ with respect to $\boldsymbol{x}_T^s$ and $\mathbf{r}$, and use Adam optimizer to solve the above optimization.

\textbf{Backdoor Detection} We detect backdoors by measuring the synthesized trigger $\boldsymbol{g}^s$'s ability to reduce sampling diversity. We introduce the Max Similarity Cluster Ratio (MSCR) to quantify similarity among embeddings $\boldsymbol{Z}^s$ of generated samples, assessing the impact of $\boldsymbol{g}^s$ to detect backdoors.

\begin{equation}
\text{MSCR}(\boldsymbol{Z})=\frac{\mid\left\{\boldsymbol{Z}_i, \boldsymbol{Z}_j \in \boldsymbol{X}: \mathcal{A}\left(\boldsymbol{Z}_i, \boldsymbol{Z}_j\right)>\kappa \right\} \mid}{|\boldsymbol{Z}|}
\label{eq:mscr}
\end{equation}
where $\mathcal{A}(\cdot)$ is the cosine similarity function and $\left|\cdot\right|$ is the set's cardinality. We consider that a model is backdoored if $\text{MSCR}\geq \kappa_d$ and set $\kappa=0.9,\ \kappa_d=0.5$ in all experiments based on empirical observations.

\subsection{Backdoor Removal (Stage 2)}

Upon detecting a backdoor, one option is to discard the model entirely, but this approach sacrifices its utility. To address this issue, we propose a structural pruning-based approach to eliminate the backdoor while preserving the model's utility. Our method utilizes the inverted trigger $\boldsymbol{g}^s$ to prune channels that are important for backdoor sampling but less important for clean sampling. Following Diff-Pruning \cite{fang2024structural}, we approximate channel importance using the Taylor expansion of the loss function.

For simplicity, we consider the backdoored diffusion model's parameter $\boldsymbol{\theta}$ as a 2-D matrix, where each row vector $\boldsymbol{\theta}_i=\left[\theta_{i 0}, \theta_{i 1}, \ldots, \theta_{i K}\right]$ contains $K$ scalar parameters. The pruning objective is to obtain a sparse parameter $\boldsymbol{\theta}^{\prime}$, minimizing the performance loss on clean data while maximizing the disruption on backdoored data. This is formalized as:

\begin{equation}
\begin{aligned}
\min_{\boldsymbol{\theta}^{\prime}} & \left|\mathcal{L}^c\left(\boldsymbol{\theta}^{\prime}, \boldsymbol{X}^c\right) - \mathcal{L}^c(\boldsymbol{\theta}, \boldsymbol{X}^c)\right|, \\
\max_{\boldsymbol{\theta}^{\prime}} & \left|\mathcal{L}^b\left(\boldsymbol{\theta}^{\prime}, \boldsymbol{X}^b\right) - \mathcal{L}^b(\boldsymbol{\theta}, \boldsymbol{X}^b)\right|, \ \text{s.t. } \left\|\boldsymbol{\theta}^{\prime}\right\|_0 \leq s
\end{aligned}
\label{eq:pruningobjective}
\end{equation}
where $\mathcal{L}^c(\cdot)$ and $\mathcal{L}^b(\cdot)$ are loss functions of clean and backdoor training respectively, $\boldsymbol{X}^c$ and $\boldsymbol{X}^b$ are clean and backdoored inputs, and $s$ is the sparsity of $\boldsymbol{\theta}^{\prime}$. $\left\|\cdot\right\|_0$ is the $L_0$ norm, which counts the number of non-zero parameters. Since the defender is unware of the attack settings, we calculate $\boldsymbol{X}^b$ using Eq. \ref{eq:backdoordiff}, where the coefficients $\hat{\alpha}_t$ and $\hat{\beta}_t$ mirror those in clean diffusion. We assume that $\lim_{t\to T} \hat{\rho}_t=1$ and $\lim_{t\to 0} \hat{\rho}_t=0$, so $\hat{\rho}_t$ is linearly interpolated from 2e-4 to 1. Due to the iterative nature of diffusion process, $\mathcal{L}^c$ and $\mathcal{L}^b$ can be decomposed into sequences $\left\{\mathcal{L}^{c}_t\right\}_{t=1}^T$ and $\left\{\mathcal{L}^{b}_t\right\}_{t=1}^T$. 

\textbf{Taylor Importance Score} Let $\mathcal{L}_{t}$ denotes either clean or backdoor training loss at timestep $t$, $\left\|\cdot\right\|^2$ the $L_2$ norm and $\boldsymbol{X}$ the inputs at $t$. Leveraging Taylor expansion \cite{molchanov2019importance}, the loss disruption can be approximated as:

\begin{equation}
\begin{aligned}
\mathcal{L}_t = \left\|\boldsymbol{\epsilon}-\boldsymbol{\epsilon}_\theta\left(\boldsymbol{X}, t\right)\right\|^2 \\
\mathcal{L}_t\left(\boldsymbol{\theta}^{\prime},\boldsymbol{X}\right) - \mathcal{L}_t(\boldsymbol{\theta},\boldsymbol{X}) & =
\nabla \mathcal{L}_t(\boldsymbol{\theta},\boldsymbol{X})\left(\boldsymbol{\theta}^{\prime}-\boldsymbol{\theta}\right) \\
& \quad + O\left(\left\|\boldsymbol{\theta}^{\prime}-\boldsymbol{\theta}\right\|^2\right)
\end{aligned}
\label{eq:taylorexpansion}
\end{equation}
Using the Taylor importance criterion, we can obtain the importance of an individual weight $\boldsymbol{\theta}_{i k}$ by setting $\boldsymbol{\theta}^{\prime}_{i k}=0$ in Eq. \ref{eq:taylorexpansion}. For structural pruning, the smallest unit of pruning is $\boldsymbol{\theta}_i$, so we adopt the aggregated Taylor importance \cite{fang2024structural} as the importance score:

\begin{equation}
\begin{aligned}
\mathcal{I}_t\left(\boldsymbol{\theta}_i, \boldsymbol{X}\right) 
&= \sum_k\left|{\mathcal{L}_t\left(\boldsymbol{\theta},\boldsymbol{X} \mid \boldsymbol{\theta}_{i k}=0 \right) - \mathcal{L}_t(\boldsymbol{\theta},\boldsymbol{X})}\right| \\
&= \sum_k\left|\boldsymbol{\theta}_{i k} \cdot \nabla_{\boldsymbol{\theta}_{i k}} \mathcal{L}_t(\boldsymbol{\theta}, \boldsymbol{X})\right|
\end{aligned}
\label{eq:It}
\end{equation}

We further analyze the contribution of different loss terms $\left\{\mathcal{L}_1, \ldots, \mathcal{L}_T\right\}$ during backdoor diffusion. We observe that in backdoor diffusion, large $t$ corresponds to large $\mathcal{L}_{b,t}$, while in clean diffusion, small $t$ corresponds to large $\mathcal{L}_{c,t}$. \cite{fang2024structural} empirically demonstrates that small $\mathcal{L}_t$ values are less informative. Thus, we introduce a threshold parameter $\mathcal{T}$ and only use timestep $t>\mathcal{T}$ for gradient accumulation, which yields the final importance score:
\begin{equation}
\mathcal{I}\left(\boldsymbol{\theta}_i, \boldsymbol{X}\right)=\sum_k\left|\boldsymbol{\theta}_{i k} \cdot \sum_{\left\{t \left\lvert\, t>\mathcal{T}\right.\right\}} \nabla_{\boldsymbol{\theta}_{i k}} \mathcal{L}_t(\boldsymbol{\theta}, \boldsymbol{X})\right|
\label{eq:I}
\end{equation}

\paragraph{Model Repair} After obtaining Taylor importance scores for substructures, we prune the top $p\%$ of channels that show the greatest difference in $\mathcal{I}\left(\boldsymbol{\theta}_i, \boldsymbol{X}^c\right)$ and $\mathcal{I}\left(\boldsymbol{\theta}_i, \boldsymbol{X}^b\right)$ to obtain $\boldsymbol{\theta}^{\prime}$. Then we fine-tune the pruned model.

Recall that the objective function of clean training is ${loss}_{c}=\left\|\boldsymbol{\epsilon}-\boldsymbol{\epsilon}_{\theta}\left(\sqrt{\bar{\alpha}_t} \boldsymbol{x}+\sqrt{1-\bar{\alpha}_t} \epsilon, t\right)\right\|^2$. However, this loss alone is inefficient because defenders have limited clean samples and computing resources. Therefore, we use the output of the unpruned model $\boldsymbol{\theta}$ as a reference, aligning the pruned model's distribution with the clean distribution obtained from training on the entire clean dataset. This reference loss can be expressed as ${loss}_{b}=\left\|\boldsymbol{\epsilon}_{\boldsymbol{\theta}^{\prime}}\left(\sqrt{\bar{\alpha}_t} \boldsymbol{x}+\sqrt{1-\bar{\alpha}_t} \boldsymbol{\epsilon}, t\right)-\epsilon_{\boldsymbol{\theta}}\left(\sqrt{\bar{\alpha}_t} \boldsymbol{x}+\sqrt{1-\bar{\alpha}_t} \epsilon, t\right)\right\|^2$, and the complete loss is ${loss}_{p}={loss}_{c}+{loss}_{b}$.

\section{Experiment}

This section outlines the experimental setup, evaluates the effectiveness of our proposed methods, and presents ablation studies. Additionally, we discuss a potential countermeasure from the attacker's perspective. The main experiments are conducted on a server equipped with Intel Golden 6240 CPUs and NVIDIA V100 GPUs.

\subsection{Experimental Setup}
\textbf{Datasets and Attacks} We use four benchmark vision datasets: MNIST ($28\times 28$) \cite{lecun1998gradient}, CIFAR-10 ($32\times 32$) \cite{krizhevsky2009learning}, CelebA ($64\times 64$) \cite{liu2015deep} and CelebA-HQ ($256\times 256$) \cite{liu2015deep}, since they are the datasets used in the addressed backdoor attack methods. Our study encompasses all badkdoor attacks designed for diffusion models (DMs) in the literature, specifically BadDiff \cite{chou2023backdoor}, TrojDiff \cite{chen2023trojdiff} and VillanDiff \cite{chou2023villandiffusion}, within the context of unconditional image generation. The evaluation spans a wide range of DMs and samplers used in these attacks, including DDPM, NCSN, LDM, DDIM, PNDM, DEIS, DPMO1, DPMO2, DPMO3, DPM++O1, DPM++O2, DPM++O3, UNIPC, and HEUN. The clean and backdoored DMs are either downloaded from Hugging Face or trained by ourselves using official codes. 

\paragraph{Baselines} As previously discussed, most existing backdoor defenses are not directly applicable for DMs. For trigger inversion, We adopt the method from Elijah \cite{an2024elijah} as the baseline, which was the only available method specifically designed for DMs at the onset of our research. To the best of our knowledge, there are no other pruning-based model repair methods for backdoored DMs. Thus, we propose and evaluate two novel pruning strategies.

\paragraph{Metrics} Following existing literature \cite{chou2023villandiffusion,an2024elijah}, we employ the following metrics: (1) \textbf{Fréchet Inception Distance} (FID) \cite{heusel2017gans} evaluates the quality of generated clean samples relative to the training dataset, with lower values indicating better utility. (2) \textbf{Detection Accuracy} (ACC) is the proportion of correctly identified clean and backdoored DMs. (3) \textbf{Attack Success Rate} (ASR) quantifies the effectiveness of the real and inverted triggers. For D2I attacks, where the target is an image, ASR is the percentage of backdoor sampling results that are sufficiently similar to the target image (i.e., the MSE with respect to the target image is below a certain threshold). For Din and Dout attacks, where target is a class of images, ASR is the percentage of backdoor sampling results that are classified as the target class by a classifier.

We use ACC and ASR to evaluate trigger inversion methods, where high ACC indicates successful distinction between clean and backdoored DMs, and high ASR shows that the synthesized trigger effectively activates the backdoor. For model repair methods, we focus on metrics changes, denoted as $\Delta \mathrm{FID}$ and $\Delta \mathrm{ASR}$, where lower values are preferable. Lower $\Delta \mathrm{FID}$ means better preservation of benign utility. Lower $\Delta \mathrm{ASR}$, calculated with the real trigger, indicates more effective backdoor removal. We provide more details about experimental settings in appendix D.

\begin{table}[htbp]
\small
\caption{Evaluation of the trigger inversion methods. Results are averaged across models of the same type. DDPM-C (resp. DDPM-A) refers to DDPM models trained on the CIFAR-10 (resp. CelebA-HQ) dataset. ODE-C and ODE-A shows the average results of ODE samplers attacked by VillanDiff. D2I means that the target distribution is an image, while Din (resp. Dout) means that the target distribution is a class of images within (resp. outside) the training dataset.}
\label{tab:triggerinversion}
  
\begin{tabular}{cccccc} 
\toprule
\multirow{2}{*}{Attack} & \multirow{2}{*}{Model} & \multicolumn{2}{c}{Elijah} & \multicolumn{2}{c}{BBTI (Ours)}\\ \cmidrule(r){3-4} \cmidrule(r){5-6}
 & & ACC $\uparrow$ & ASR $\uparrow$ & ACC $\uparrow$ & ASR $\uparrow$ \\

\midrule \multirow{2}{*}{BadDiff}
& DDPM-C & 1.00 & 0.56 & 1.00 & 0.90 \\
& DDPM-A & 1.00 & 1.00 & 1.00 & 1.00 \\

\midrule \multirow{4}{*}{VillanDiff} & ODE-C & 0.62 & 0.09 & 0.97 & 0.83 \\
 & ODE-A & 1.00 & 0.00 & 1.00 & 0.95 \\
 & NCSN-C & 1.00 & 0.12 & 1.00 & 0.88 \\
 & LDM-A & 0.50 & 0.00 & 1.00 & 1.00 \\

 \midrule \multirow{3}{*}{TrojDiff} & D2I & 1.00 & 1.00 & 1.00 & 1.00\\
 & Din & - & - & 1.00 & 0.71 \\
  & Dout & - & - & 1.00 & 0.73 \\
\bottomrule
\end{tabular}
\end{table}

\subsection{Evaluation of Trigger Inversion}

Tab. \ref{tab:triggerinversion} compares the performance of trigger invertion method in Elijah and our proposed method BBTI. Following the original paper, we use clean and backdoored models to train a random forest as the backdoor detector of Elijah. BBTI use MSCR in Eq. \ref{eq:mscr} for backdoor detection. 

While Elijah generally succeeds in distinguishing between clean and backdoored diffusion models, it is less effective against models attacked by VillanDiff. In cases where detection fails, the metrics used by Elijah, average total variance loss and uniformity loss, show minimal differences between clean and backdoored models, leading to diminished detection accuracy. Additionally, the triggers inverted by Elijah exhibit low ASR, indicating their limited effectiveness in activating backdoors, which is a critical limitation in practical applications. In real-world scenarios, the lack of visual evidence corresponding to the target distribution can undermine confidence in detection, even if metrics indicate the presence of a backdoor. In contrast, our proposed method, BBTI, demonstrates robust performance across all evaluated attacks, achieving high ACC in backdoor detection and generating triggers that align well with target distributions (high ASR). Moreover, our approach does not rely on clean or backdoored models for training, making it more resource-efficient.

\begin{table}[htbp]
\small
\centering
\setlength{\tabcolsep}{3.5pt}
\caption{Results of two pruning-based defense strategies. We report the average performance across models of the same type. For TrojDiff, we report the average results for models trained on the same dataset. “C” always denotes the CIFAR-10 dataset, while “A” denotes the CelebA dataset for TrojDiff and CelebA-HQ dataset for BadDiff and VillanDiff.}
\label{tab:backdoorremoval}
\begin{tabular}{cccccc}
\hline
\multirow{2}{*}{Attack} & \multirow{2}{*}{Model} & \multicolumn{2}{c}{Diff-Pruning} & \multicolumn{2}{c}{Diff-Cleanse (Ours)}\\ \cmidrule(r){3-4} \cmidrule(r){5-6}
 & & $\Delta \mathrm{FID}\downarrow$ & $\Delta \mathrm{ASR}\downarrow$ & $\Delta \mathrm{FID}\downarrow$ & $\Delta \mathrm{ASR}\downarrow$\\
 \midrule \multirow{2}{*}{BadDiff}
 & DDPM-C & 8.37 & -1.00 & \textbf{3.50} & -1.00 \\
 & DDPM-A & 8.03 & -1.00 & \textbf{3.14} & -1.00 \\
 \midrule \multirow{4}{*}{VillanDiff}
 & ODE-C & 4.38 & -0.96 & \textbf{2.76} & -0.96 \\
 & ODE-A & 6.82 & -0.93 & \textbf{3.99} & -0.93 \\
 & NCSN-C & 7.89 & -0.94 & \textbf{3.62} & -0.94 \\
 & LDM-A & 6.45 & -0.96 & \textbf{3.83} & -0.96 \\

 \midrule \multirow{4}{*}{TrojDiff} & DDPM-C & 6.44 & -1.00 & \textbf{4.05} & -1.00 \\
 & DDIM-C & 8.49 & -0.91 & \textbf{2.84} & -0.91 \\
 & DDPM-A & 5.51 & -0.89 & \textbf{3.16} & -0.89 \\
 & DDIM-A & 7.02 & -0.78 & \textbf{3.95} & -0.78 \\

\hline
\end{tabular}
\end{table}

\subsection{Evaluation of Backdoor Removal}

Fig. \ref{tab:backdoorremoval} compares two pruning-based backdoor removal strategies, Diff-Pruning \cite{fang2024structural} and our proposed Diff-Cleanse. $\Delta\text{ASR}$ is calculated using real triggers. Following the original paper, Diff-Pruning eliminates channels that contribute the least to clean sampling, while Diff-cleanse prunes based on differences in channel sensitivity to the inverted trigger. By pruning a sufficient portion of the model's substructure, both methods can fully remove the backdoor. However, Diff-Pruning requires pruning $7\%\sim 13\%$ of channels to achieve this, whereas Diff-Cleanse only needs to prune $1\%\sim 2\%$, thereby preserving more of the model's benign utility, as indicated by lower $\Delta\text{FID}$ scores. In contrast, after pruning $2\%$ of channels, Diff-Pruning leaves both the model's benign performance and backdoor largely intact. Please check appendix D for detailed settings and appendix F for visual examples. 

\begin{table}[htbp]
\centering
\setlength{\tabcolsep}{3.5pt}
\caption{The average backdoor detection performance of BBTI when using only $\ell_s$ or $\ell_s+\ell_e$ against three attacks. }
\label{tab:effectofle}
\begin{tabular}{ccccc}
\hline
\multirow{2}{*}{Attack} & \multicolumn{2}{c}{$\ell_s$} & \multicolumn{2}{c}{$\ell_s+\ell_e$}\\ \cmidrule(r){2-3} \cmidrule(r){4-5}
 & ACC & ASR & ACC & ASR \\
 \midrule BadDiff& 0.65 & 0.61 & \textbf{1.00} & 0.92 \\
 \midrule VillanDiff & 0.51 & 0.44 & \textbf{0.97} & 0.91 \\
 \midrule TrojDiff & 0.75 & 0.61 & \textbf{1.00} & 0.84 \\ 
\hline
\end{tabular}
\end{table}

\subsection{Ablation study}

In this section, we study the effect of components and hyperparameters to the performance of our proposed defense. We provide visual examples in Appendix F.

\paragraph{Effect of $\ell_e$.} Tab. \ref{tab:effectofle} campares the performance of our trigger inversion method with and without $\ell_e$. The upper bound $\kappa_{+}=8.5$ and lower bound $\kappa_{-}=0.1$ used in our experiments are relatively loose compared to the real distribution, as shown in Fig. \ref{fig:entropyboxplot}. Even when these bounds are tightened to $\kappa_{+}=7.5,\ \kappa_{-}=1.0$, the algorithm’s performance remains largely unaffected. However, omitting $\ell_e$ leads to a significant decline in performance. In failed cases, backdoor sampling with the inverted trigger would generate produce similar pure color or noise images.

\begin{table}[htbp]
\small
\centering
\setlength{\tabcolsep}{3.5pt}
\caption{Fine-tuning results of Diff-Cleanse using general objective ${loss}_c$ and our proposed ${loss}_p$.}
\label{tab:lossc_p}
\begin{tabular}{cccccc} 
  \toprule
\multirow{2}{*}{Attack} & \multirow{2}{*}{Model} & \multicolumn{2}{c}{${loss}_c$} & \multicolumn{2}{c}{$loss_p$}\\ \cmidrule(r){3-4} \cmidrule(r){5-6}
 & &$\Delta \mathrm{FID}\downarrow$ & $\Delta \mathrm{ASR}\downarrow$ & $\Delta \mathrm{FID}\downarrow$ & $\Delta \mathrm{ASR}\downarrow$ \\

  \midrule \multirow{2}{*}{BadDiff}
& DDPM-C & 5.81. & -1.00  & \textbf{3.96} & -1.00 \\
& DDPM-A & 5.72 & -1.00 & \textbf{4.13} & -1.00 \\

\midrule 
\multirow{2}{*}{VillanDiff} & ODE-C & 2.62 & -0.98 & \textbf{1.89} & -0.98 \\
 & ODE-A & 4.64 & -0.94 & \textbf{2.97} & -0.94 \\
 & NCSN-C & 6.25 & -0.96 & \textbf{4.17} & -0.96 \\
 & LDM-A & 5.41 & -0.96 & \textbf{3.03} & -0.96\\

\midrule \multirow{2}{*}{TrojDiff} & DDIM-C & 5.39 & -0.91 & \textbf{3.15} & -0.91 \\
 & DDIM-A & 6.34 & -0.82 & \textbf{3.90} & -0.82 \\
\bottomrule
\end{tabular}
\end{table}

\begin{figure}  
\centering
\includegraphics[width=0.9\linewidth]{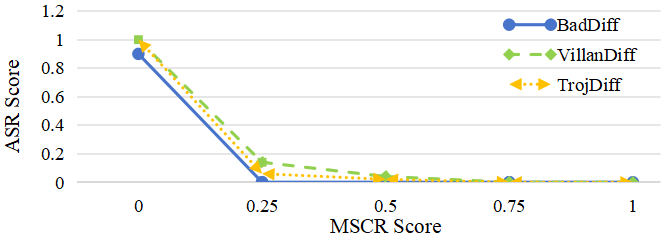}
\caption{ASR scores for pruned diffusion models with inverted triggers of varying effectiveness. A lower ASR means a more thorough removal of the backdoor.}
\label{fig:triggereffect}
\end{figure}

\paragraph{Effect of ${loss}_p$.}
Tab. \ref{tab:lossc_p} compares the fine-tuning results of Diff-Cleanse using ${loss}_c$ and ${loss}_p$. Due to the limited computing resources, we test part of models from each attack method and report the average results. We observe that ${loss}_p$ always lead to lower FID scores, indicating that using the prediction of the unpruned model as a reference can effectively improve fine-tuning performance.

\paragraph{Effect of Trigger Effectiveness.} In Fig. \ref{fig:triggereffect}, the MSCR score is calculated with the inverted trigger and the ASR score with the real trigger. Even when the synthesized trigger is less effective in activating the backdoor (low MSCR), the pruning method still successfully eliminates the backdoor (ASR close to zero), demonstrating the robustness of our approach to the effectiveness of inverted triggers.

\subsection{Adaptive Attack}

After realizing the criteria of Diff-Cleanse, the attacker may train more concealed backdoor diffusion models, where the difference in activation values of clean noise and backdoored noise is reduced. To simulate this situation, we design adaptive versions of BadDiffusion and VillanDiffusion. Specifically, we calculate the Euclidean distance between the activation of clean noise and backdoored noise at the same timestep, and use it as a weighted term in the loss. The detailed method and experimental results are in appendix G.

\section{Conclusion}
\label{sec:conclusion}
In this paper, we present Diff-Cleanse, a novel two-stage backdoor defense framework. Extensive experiments show that Diff-Cleanse has a detection accuracy close to 100\% and can completely remove backdoors without significantly affecting the model's benign performance.


\clearpage  
\appendix
\section{A Main Branches of Diffusion Models}
In this section, we introduce three major branches of diffusion models, Denoising Diffusion Probabilistic Model (DDPM) \cite{ho2020denoising}, score-based models \cite{NEURIPS2019_3001ef25,song2020score} and latent diffusion model (LDM) \cite{rombach2022high}, as well as advanced samplers.

\textbf{DDPM} is a representative diffusion model. Let $\boldsymbol{x}_t$ denotes the input of the model at timestep $t$. DDPM assumes the prior distribution $\tau(\boldsymbol{x})\sim \mathcal{N}(0,\mathbf{I})$ and defines the Markov diffusion process $q\left(\boldsymbol{x}_t \mid \boldsymbol{x}_{t-1}\right)=N\left(\boldsymbol{x}_t; \sqrt{1-\beta_t} \boldsymbol{x}_{t-1},\beta_t\mathbf{I}\right)$, where $\left\{\beta_t\right\}_{t=1}^T$ is a predefined monotonically increasing  variance squence. Usually, $\beta_1$ is close to 0 and $\beta_T=1$. Let $\alpha_t=1-\beta_t$ and $\bar{\alpha}_i=\prod_{t=1}^i \alpha_t$. Given $\boldsymbol{x}_0 \sim q(\boldsymbol{x}_0), t \sim \operatorname{Uniform}(\{1, \ldots, T\})$ and $\boldsymbol{\epsilon} \sim \mathcal{N}(0, I)$, the diffusion model is trained to minimize the loss $\| \boldsymbol{\epsilon}-\boldsymbol{\epsilon}_\theta\left(\sqrt{\bar{\alpha}_t} x_0+\right.\left.\sqrt{1-\bar{\alpha}_t} \boldsymbol{\epsilon}, t\right) \|^2$, thereby learning to predict the added noise given the $\boldsymbol{x}_t$ at timestep $t$. $q(\boldsymbol{x}_0)$ is the training dataset distribution. During sampling, given $\boldsymbol{x}_T \sim \mathcal{N}(0, I)$, DDPM samples from $p_{\theta}\left(\boldsymbol{x}_{t-1} \mid \boldsymbol{x}_t\right)=\mathcal{N}\left(\boldsymbol{x}_{t-1} ; \mu_{\theta}\left(\boldsymbol{x}_t\right), \beta_{\theta}\left(\boldsymbol{x}_t\right)\right)$ from $t=T$ to $t=1$, where $\mu_\theta\left(\boldsymbol{x}_t\right)=\frac{1}{\sqrt{\alpha_t}}\left(\boldsymbol{x}_t-\frac{1-\alpha_t}{\sqrt{1-\bar{\alpha}_t}} \boldsymbol{\epsilon}_\theta\left(\boldsymbol{x}_t, t\right)\right)$ and $\beta_\theta\left(\boldsymbol{x}_t\right)=\frac{\left(1-\bar{\alpha}_{t-1}\right) \beta_t}{1-\bar{\alpha}_t}$. 

\textbf{Score-based models} are trained to estimate the score function, which represents the gradient of the log-density of the data. \cite{song2020score} propose a unified Stochastic Diffusion Equation (SDE)-based framework, which characterizes the diffusion process by the following SDE:

\begin{equation}
d \boldsymbol{x}_t=f\left(\boldsymbol{x}_t, t\right) d t+g(t) d \mathbf{w}
\end{equation}
where $t \in[0, T]$ and $f\left(\boldsymbol{x}_t, t\right), g(t)$ are the drift and diffusion coefficients, respectively. According to \cite{anderson1982reverse}, the denoising process corresponds to a reversed SDE:

\begin{equation}
d \boldsymbol{x}_t=\left[f\left(\boldsymbol{x}_t, t\right)-g(t)^2 \nabla_{\boldsymbol{x}} \log p_t\left(\boldsymbol{x}_t\right)\right] d t+g(t) d \mathbf{w}
\label{eq:reversesde}
\end{equation}

Given $f\left(\boldsymbol{x}_t, t\right)$ and $g(t)$, only $\nabla_{\boldsymbol{x}} \log p_t\left(\boldsymbol{x}_t\right)$, the score of $\boldsymbol{x}_t$, is unknown in Eq. \ref{eq:reversesde}. However, we can train the model $s_\theta\left(\boldsymbol{x}_t, t\right)$ with loss $\left\|s_\theta\left(\boldsymbol{x}_t, t\right)-\nabla_{\boldsymbol{x}} \log p_t\left(\boldsymbol{x}_t\right)\right\|^2$ to fit it. Using initial value $\boldsymbol{x}_{T_{\max}} \sim \mathcal{N}(0, \sigma^2 I)$ and solving Eq. \ref{eq:reversesde}, the model can generate high-quality and diverse images.

\textbf{LDM} can be regarded as DDPM operating in the compressed latent space of a pre-trained encoder $\mathcal{E}$. The diffusion chain of LDM does not produce an image $\boldsymbol{x}_0$, but produces a latent vector $\boldsymbol{z}_0$, which could be reconstruct to images by a decoder $\mathcal{D}$.

\textbf{Advanced Samplers} The default samplers for diffusion models are relatively slow compared to other generative models such as GANs \cite{NEURIPS2019_3001ef25,ho2020denoising}, prompting the development of advanced samplers to accelerate the sampling.
DDIM \cite{song2020denoising} introduces an abbreviated reverse process, reducing the steps from 1000 to as few as 50, based on a generalized non-Markovian forward chain derived from DDPM. Additionally, some researchers formulate the reversed diffusion process as Ordinary Differential Equations (ODEs) and employ higher-order approximations, enabling diffusion models to generate high-quality samples in fewer steps \cite{song2020score,lu2022dpm,lu2022dpmplus,zhao2024unipc,karras2022elucidating,zhang2022fast}.

\section{B Parameter Instantiation for Backdoor Attacks on Diffusion Models}
Recall that the unified form of backdoor diffusion is $\boldsymbol{x}_t^{b}=\hat{\alpha}_t \boldsymbol{x}_0^{b}+\hat{\beta}_t \boldsymbol{\epsilon}_t+\hat{\rho}_t \mathbf{r}$, as shown in \ref{eq:backdoordiff}, this section describes the forward diffusion process of existing backdoor attacks. In this section, $\boldsymbol{x}_t$ denotes the input of the diffusion model at timestep $t$, $\boldsymbol{\epsilon} \sim \mathcal{N}(0, \mathbf{I})$ denotes the random noise and $\mathbf{r}$ denotes the value of the trigger. The superscript $c$ and $b$ denotes the data is used in clean and backdoor diffusion respectively.

\textbf{Clean Diffusion Process} Utilizing the diffusion process $q\left(\boldsymbol{x}_t \mid \boldsymbol{x}_{t-1}\right)$ in appendix A and the Bayes Rule, given $\boldsymbol{x}_0$, the forward diffusion process is as follows:  

\begin{equation}
\boldsymbol{x}_t^c=\sqrt{\bar{\alpha}_t} \boldsymbol{x}_0^c+\sqrt{1-\bar{\alpha}_t} \boldsymbol{\epsilon}
\end{equation}

\textbf{BadDiff} sets $\hat{\alpha}_t=\sqrt{\bar{\alpha}_t}$, $\hat{\beta}_t=\sqrt{1-\bar{\alpha}_t}$ and $\hat{\rho}_t=1-\sqrt{\bar{\alpha}_t}$, so the backdoor diffusion is as follows:

\begin{equation}
\boldsymbol{x}_t^b=\sqrt{\bar{\alpha}_t} \boldsymbol{x}_0^b+\sqrt{1-\bar{\alpha}_t} \boldsymbol{\epsilon}+(1-\sqrt{\bar{\alpha}_t})\mathbf{r} 
\end{equation}

\textbf{VillanDiff} refers to $\hat{\alpha}_t$ and $\hat{\beta}_t$ as the content schedulers and noise schedulers respectively, and set their values in two cases. For DDPM and LDM, the forward process is the same as in BadDiff, but the authors modify the loss objective to enable the method to attack advanced samplers. For score-based diffusion models, $\hat{\alpha}_t=1$, $\hat{\beta}_t$ increases with $t$, and $\hat{\rho}_t=\hat{\beta}_t$.

\textbf{TrojDiff} sets $\hat{\alpha}_t=\sqrt{\bar{\alpha}_t}$, $\hat{\beta}_t=\sqrt{1-\bar{\alpha}_t} \gamma$ and $\hat{\rho}_t=\sqrt{1-\bar{\alpha}_t}$, 
so the backdoor diffusion is as follows:
\begin{equation}
\boldsymbol{x}_t^b=\sqrt{\bar{\alpha}_t} \boldsymbol{x}_0^b+\sqrt{1-\bar{\alpha}_t} \gamma \epsilon_t+\sqrt{1-\bar{\alpha}_t}\mathbf{r}
\end{equation}
where $\gamma$ is the hyperparameter used to control the range of $\boldsymbol{x}_t$. TrojDiff limits $\boldsymbol{x}_t$ to $[-1,1]$, while the other two methods do not have this limitation.

\begin{algorithm}
\caption{Trigger Inversion}
\label{alg:triggerinversion}
\begin{algorithmic}[1]
\REQUIRE diffusion model $M$, size of the batch $bz$, dimension of the input $dim$, encoder $\mathcal{E}$
\ENSURE inverted trigger $\mathbf{r}$, backdoor indicater $flag$ 
\STATE Initialize:
\STATE Generate random noise $\boldsymbol{\epsilon} \sim \mathcal{N}(0,1)$ with shape $(bz, dim)$
\STATE Generate trigger $\mathbf{r} \sim \mathcal{N}(0,1)$ with shape $(1, dim)$
\STATE $flag=0$
\REPEAT
    \STATE Extend $\mathbf{r}$ to shape $(bz, dim)$
    \STATE $\boldsymbol{X}_T = \boldsymbol{\epsilon} + \mathbf{r}$
    \STATE $\boldsymbol{X}_0 = \text{backdoor sampling}(\boldsymbol{X}_T)$
    \STATE Calculate $\text{MSCR}(\boldsymbol{X}_0)$ $\hfill\triangleright$ \ref{eq:mscr}
    \IF{$\text{MSCR}(\boldsymbol{X}_0)>0.5$}
    \STATE $flag=1$
    \STATE Break
    \ENDIF

    \STATE $Z=\mathcal{E}(\boldsymbol{X}_0)$
    \STATE Calculate $\ell\left(\mathbf{r}\right)$ and update $\mathbf{r}$ $\hfill\triangleright$ \ref{eq:l(g)}
    
\UNTIL{Reach the max number of iterations.}

\RETURN $\mathbf{r}, \ flag$
\end{algorithmic}
\end{algorithm}

\section{C Pseudocode of Our Algorithms}
\subsection{Trigger Inversion}
Algorithm 1 summarizes our black box trigger inversion
and backdoor detection method. Line 2 samples a batch of clean noise from standard Gaussian noise. Lines 3-4 initializes the trigger $\boldsymbol{g}$ and the backdoor indicater $flag$. Lines 5-16 optimize $\boldsymbol{g}$ based on our proposed objective, using Adam optimizer with a learning rate of 0.1 and a maximum of 100 iterations in all experiments. Specifically, lines 6-8 implement backdoor sampling with the inverted trigger. Line 9 calculates our proposed metric MSCR with $\kappa=0.9$ for D2I attack and $\kappa=0.75$ for Din and Dout attacks. Lines 10-13 determine whether the presence of a backdoor. Lines 14-15 calculate our proposed optimization objective $\ell(\boldsymbol{g})$ and update the trigger $\mathbf{r}$.
Line 17 returns the inverted trigger and the backdoor indicator.

\subsection{Backdoor Removal}
Algorithm 2 outlines the procedure of removing the backdoor and restoring the benign utility of the pruned model. Line 1 randomly chooses a batch of clean samples from the clean dataset. Line 2 samples a clean noise. Lines 3-11 calculate the loss of $\boldsymbol{\theta}$ against clean inputs and backdoored inputs, as well as the gradients with respect to the substructures $\boldsymbol{\theta}_{ik}$. Lines 12-17 calculate the aggregated Taylor importance scores as shown in Eq. \ref{eq:I}. Lines 18-20 remove parameters based on their importance scores and finetune the pruned model with our proposed objective ${loss}_p$. Line 21 returns the purified model $\boldsymbol{\theta}^\prime$.

\begin{algorithm}[t]
\caption{Structural Pruning-based Backdoor Removal}
\label{alg:structuralpruning}
\begin{algorithmic}[1]
\REQUIRE A diffusion model $\boldsymbol{\theta}=\left[\boldsymbol{\theta}_0, \dots, \boldsymbol{\theta}_M\right]$, an inverted trigger $\mathbf{r}$, clean dataset $\boldsymbol{X}$, gradients accumulation threshold $\mathcal{T}$, a pruning ratio $p\%$, diffusion coefficient sequences $\left\{\hat{\alpha}_t\right\}, \left\{\hat{\beta}_t\right\},\left\{\hat{\rho}_t\right\}$
\ENSURE The pruned diffusion model $\boldsymbol{\theta}^{\prime}$ 
\STATE $\boldsymbol{x}^c_0=\text{mini-batch}(\boldsymbol{X})$
\STATE $\boldsymbol{\epsilon} \sim \mathcal{N}(0,1)$
\STATE $\textcolor{gray}{\triangleright \text{ Accumulating gradients over partial steps with } \mathcal{T}}$

\FOR {$t\text{ in } \left[\mathcal{T}, \mathcal{T}+1, \dots, T\right]$}
    \STATE $\boldsymbol{x}^c_t=\hat{\alpha}_t \boldsymbol{x}^c_0+\hat{\beta}_t\boldsymbol{\epsilon}$
    \STATE $\mathcal{L}^{c}_{t}(\boldsymbol{\theta}, \boldsymbol{x}^c_t)=\left\|\boldsymbol{\epsilon}-\boldsymbol{\epsilon}_{\boldsymbol{\theta}}\left(\boldsymbol{x}^c_t, t\right)\right\|^2$
    
    \STATE $\nabla_{\boldsymbol{\theta}_{i k}} \mathcal{L}^c_t(\boldsymbol{\theta},\boldsymbol{x}^c_t)=\text {back-propagation}\left(\mathcal{L}^c_t(\boldsymbol{\theta},\boldsymbol{x}^c_t)\right)$

    \STATE $\boldsymbol{x}^b_t=\hat{\alpha}_t \boldsymbol{x}^c_0+\hat{\beta}_t\boldsymbol{\epsilon}+\hat{\rho}_t\mathbf{r}$ $\hfill\triangleright$ \ref{eq:backdoordiff}

    \STATE $\mathcal{L}^{b}_{t}(\boldsymbol{\theta},\boldsymbol{x}^b_t)=\left\|\boldsymbol{\epsilon}-\boldsymbol{\epsilon}_{\boldsymbol{\theta}}\left(\boldsymbol{x}_t^{b}, t\right)\right\|^2$
    
    \STATE $\nabla_{\boldsymbol{\theta}_{i k}} \mathcal{L}^b_t(\boldsymbol{\theta},\boldsymbol{x}^b_t)=\text {back-propagation}\left(\mathcal{L}^b_t(\boldsymbol{\theta},\boldsymbol{x}^b_t)\right)$
\ENDFOR    

\STATE $\textcolor{gray}{\triangleright \text{ Estimating the importance of substructure } \theta_i}$

\FOR {$i\text{ in } \left[0, 1, \dots, M\right]$}

    \STATE $\mathcal{I}^c\left(\boldsymbol{\theta}_i\right)=\sum_k\left|\boldsymbol{\theta}_{i k} \cdot \sum_{t=\mathcal{T}}^T \nabla_{\boldsymbol{\theta}_{i k}} \mathcal{L}_t^c(\boldsymbol{\theta})\right|$

    \STATE $\mathcal{I}^b\left(\boldsymbol{\theta}_i\right)=\sum_k\left|\boldsymbol{\theta}_{i k} \cdot \sum_{t=\mathcal{T}}^T \nabla_{\boldsymbol{\theta}_{i k}} \mathcal{L}_t^b(\boldsymbol{\theta})\right|$

    \STATE $\mathcal{I}\left(\boldsymbol{\theta}_i\right)=\mathcal{I}^b\left(\boldsymbol{\theta}_i\right)-\mathcal{I}^c\left(\boldsymbol{\theta}_i\right)$
\ENDFOR

\STATE $\textcolor{gray}{\triangleright \text{ Pruning and finetuning}}$

\STATE $\text { Remove } p\% \text{ channels in each layer to obtain } \boldsymbol{\theta}^{\prime}$

\STATE $\text { Finetune the pruned model } \boldsymbol{\theta}^{\prime} \text { on } X\text { with } {loss}_p$ 

\RETURN $\boldsymbol{\theta}^{\prime}$
\end{algorithmic}
\end{algorithm}

\section{D Detailed Configurations for Experiments}

\subsection{Runs of Algorithms.} 
Due to the limitation of computing resources, for each model in the 177 clean and 196 backdoored models, we run algorithm \ref{alg:triggerinversion} and algorithm \ref{alg:structuralpruning} once.  Given the number of models we evaluated, we believe the results should be reliable.

\subsection{Metrics}
In this section, we provide a detailed rationale for the selection of metrics and explain the methods used for their calculation.
\paragraph{Fréchet Inception Distance (FID)} is a widely used metric for evaluating the quality of generated images. It measures the distance between the feature distributions of generated images and real images, computed using a pre-trained Inception network. Lower FID scores indicate closer alignment between the distributions, suggesting greater similarity to real data, and usually higher image quality. 

\paragraph{Accuracy (ACC)} is employed to assess the performance of the backdoor detection method. The detection task is modeled as a binary classification problem, where ACC is computed as the ratio of correctly classified models to the total number of models tested. This metric offers a clear and direct measure of the algorithm's effectiveness in accurately identifying the presence or absence of backdoors in the models.

\paragraph{Attack Success Rate (ASR)} is defined as the proportion of backdoor sampling results that align with the target distribution. ASR is calculated as in Eq. \ref{eq:asr}.

\begin{equation}
\label{eq:asr}
\text{ASR}= \begin{cases}
\frac{\sum_i \mathbb{I}\left(\mathbf{d}\left(\boldsymbol{x}_i, \text { target }\right)<\kappa_m \right)}{|\boldsymbol{X}|},
 & \text {for D2I attack} \\ 
 \frac{\sum_i \mathbb{I}\left(C\left(\boldsymbol{x}_i\right)=\hat{y} \right)}{|\boldsymbol{X}|}, & \text {for Din and Dout Attacks}\end{cases}
\end{equation}
where $\mathbf{d}(\cdot)$ calculates the MSE distance and $\kappa_m$ is a threshold, set to 1e-3 in our experiments. The indicator function $\mathbb{I}$ returns 1 if the condition inside the parentheses is satisfied, and 0 otherwise. $\boldsymbol{x}_i$ is the i-th sample of backdoor sampling results $\boldsymbol{X}$ and $|\cdot|$ is the cardinality of the set. $C$ is a classifier trained using the target dataset. Specifically, for CIFAR-10 and MNIST dataset, we download corresponding classifiers from HuggingFace (aaraki/vit-base-patch16-224-in21k-finetuned-cifar10 and farleyknight-org-username/vit-base-mnist), whose classification accuracy scores are 0.9788 and 0.9948 on the evaluation set. Besides, we train a ResNet18 of 79.65\% testing accuracy on CelebA. Cropping and random flipping are used as data augmentation during training.

ASR is applied for two purposes depending on the type of trigger used: (1) for inverted triggers, ASR measures the trigger's ability to activate the backdoor, reflecting its effectiveness. (2) for real triggers, ASR evaluates the model's capability to generate target images, i.e., the effectiveness of the backdoor.

To measure the utility of the model, we sample 10K images from clean sampling process to calculate the FID score. For the ASR score, we sample 1K images from backdoor sampling process, since we find this number is sufficient to demonstrate the effectiveness of the trigger or backdoor. For BadDiff, we use the DDPM sampler. For Villandiff and Trojdiff that can be applied to multiple samplers, we calculate FID and ASR separately for each sampler and report the average results across all samplers.

\subsection{Detailed Configuration for Attacks}
We consider three existing backdoor attacks for diffusion models, BadDiff, VillanDiff and TrojDiff, and use all trigger-target pairs from the original paper, as shown in Tab. \ref{tab:triggerandtarget-baddiff} and Tab. \ref{tab:triggerandtarget-trojdiff}. 

For the BadDiff and VillanDiff, we employ varying poison rates (0.1, 0.3, 0.5, 0.7, 0.9) and triggers of different sizes ($8\times 8$, $11\times 11$, $14\times 14$, $18\times 18$), to train the models. For TrojDiff, we implement the Din, Dout and D2I attack described in the paper and train models with the same hyperparameters as those used in the original paper.

\begin{table*}[htbp]
\centering
\caption{Patterns of triggers and targets for BadDiff and VillanDiff. Each image of CIFAR10 (resp. CelebA-HQ) is $32\times 32$ (resp. $256\times 256$) pixels. The target patterns, “Trigger” and “Shift”, are derived from the same pattern as the trigger. However, the “Trigger” target maintains the same position as the original trigger, while the “Shift” target is displaced to the upper left. The “Stop Sign” is used as an example to visualize both the “Trigger” and “Shift” target patterns.}
\begin{tabular}{c|c|c|c|c|c|c|c|c}
\hline \multicolumn{7}{c|}{CIFAR-10 ($32\times 32$)} & \multicolumn{2}{c}{CelebA-HQ ($256\times 256$)}\\
\midrule \multicolumn{2}{c|}{Triggers} & \multicolumn{5}{c} {Targets} & \multicolumn{1}{|c|}{Triggers} & Targets\\
\midrule Grey Box & Stop Sign & Trigger & Shift & Corner & Shoe & Hat & Glasses & Cat \\
\begin{minipage}[b]{0.15\columnwidth}
		\centering
		\raisebox{-.5\height}{\includegraphics[width=25pt]{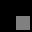}}
	\end{minipage} 
 & \begin{minipage}[b]{0.15\columnwidth}
		\centering
		\raisebox{-.5\height}{\includegraphics[width=25pt]{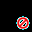}}
	\end{minipage} 
 & \begin{minipage}[b]{0.15\columnwidth}
		\centering
		\raisebox{-.5\height}{\includegraphics[width=25pt]{figures/stopsign14.png}}
	\end{minipage} 
 & \begin{minipage}[b]{0.15\columnwidth}
		\centering
		\raisebox{-.5\height}{\includegraphics[width=25pt]{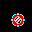}}
	\end{minipage} 
 & \begin{minipage}[b]{0.15\columnwidth}
		\centering
		\raisebox{-.5\height}{\includegraphics[width=25pt]{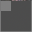}}
	\end{minipage} 
 & \begin{minipage}[b]{0.15\columnwidth}
		\centering
		\raisebox{-.5\height}{\includegraphics[width=25pt]{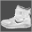}}
	\end{minipage} 
& \begin{minipage}[b]{0.15\columnwidth}
		\centering
		\raisebox{-.5\height}{\includegraphics[width=25pt]{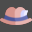}}
	\end{minipage}
& \begin{minipage}[b]{0.15\columnwidth}
		\centering
		\raisebox{-.5\height}{\includegraphics[width=25pt]{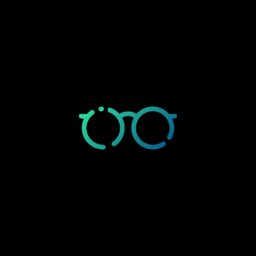}}
	\end{minipage}
& \begin{minipage}[b]{0.15\columnwidth}
		\centering
		\raisebox{-.5\height}{\includegraphics[width=25pt]{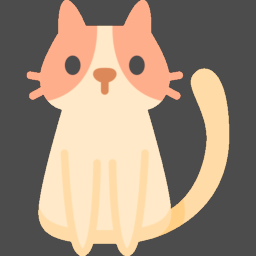}}
	\end{minipage}
\end{tabular}
\label{tab:triggerandtarget-baddiff}
\end{table*}

\begin{table*}[htbp]
\centering
\caption{Patterns of triggers and targets for TrojDiff. We use two types of triggers for all three attacks. The blend-based trigger mixes the trigger image with all the input, while the patch-based trigger only modifies the area corresponding to the white block. For Din attack, We use class “horse” for models trained on CIFAR-10 and "Heavy Makeup, Mouth Slightly Open, and Smiling" for models trained on CelebA. For Dout attack, we use digital number “7” in MNIST as the target distribution. We use “Mickey” as the target of D2I attack. }
\begin{tabular}{c|c|c|c|c}
\hline \multicolumn{2}{c|}{Triggers} & \multicolumn{3}{c}{Targets}\\
\midrule Blend-based & Patch-based & Din Attack & Dout Attack & D2I Attack \\
\begin{minipage}[b]{0.25\columnwidth}
		\centering
		\raisebox{-.5\height}{\includegraphics[width=55pt]{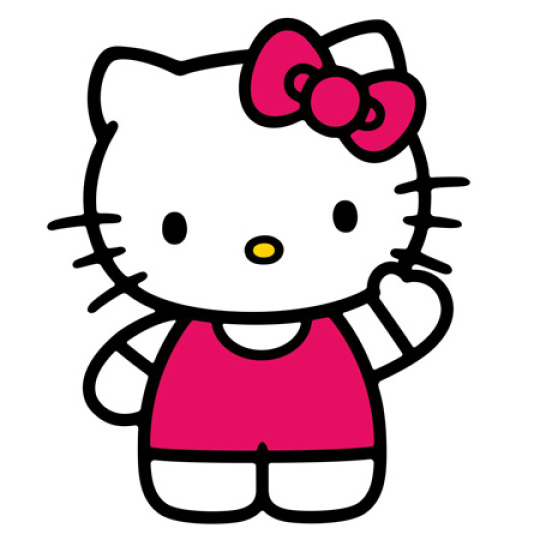}}
	\end{minipage} 
 & \begin{minipage}[b]{0.25\columnwidth}
		\centering
		\raisebox{-.5\height}{\includegraphics[width=55pt]{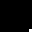}}
	\end{minipage} 
 & \begin{minipage}[b]{0.25\columnwidth}
		\centering
		\raisebox{-.5\height}{\includegraphics[width=55pt]{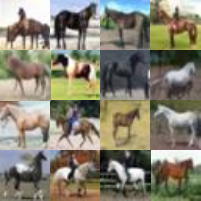}}
	\end{minipage} 
 & \begin{minipage}[b]{0.25\columnwidth}
		\centering
		\raisebox{-.5\height}{\includegraphics[width=55pt]{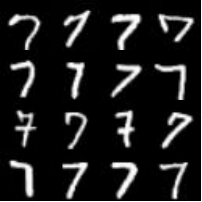}}
	\end{minipage}
 & \begin{minipage}[b]{0.25\columnwidth}
		\centering
		\raisebox{-.5\height}{\includegraphics[width=55pt]{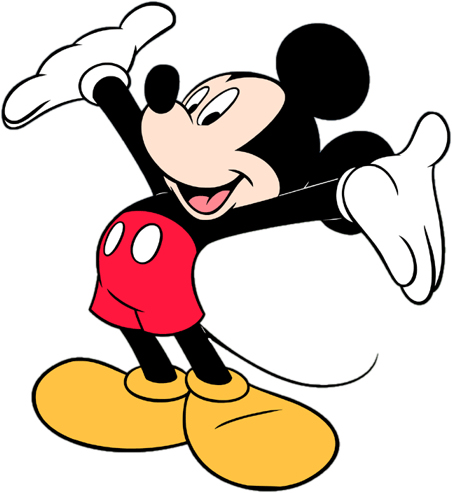}}
	\end{minipage} 
\end{tabular}
\label{tab:triggerandtarget-trojdiff}
\end{table*}

\subsection{Detailed Configuration for Defenses}
For Elijah, we optimize the trigger using the hyperparameters in the original paper \cite{an2024elijah} and train a random forest classifier as the backdoor indicator, since it is the more powerful method in the original paper. For our proposed trigger inversion method BBTI, we use $\kappa_{+}=8.5, \kappa_{-}=0.1,\kappa=0.9,\kappa_d=0.5$. We use “ViT-B/32” of Clip as the encoder to transform images into embeddings. When backdoor sampling with inverted trigger $\boldsymbol{g}^s$, we use DDPM sampler for BadDiff and DDIM sampler for VillanDiff and TrojDiff. Ideally, a larger batch size will give us a better approximation of the expectation in Eq. \ref{eq:l(g)} and Eq. \ref{eq:mscr}. We set the batch size to 100 for DMs with $3\times 32\times 32$ space, 50 for $3\times 64\times 64$ space, and 20 for $3\times 256\times 256$ space because of GPU memory limitation.

For our proposed backdoor removal method, we set $\tau$ to 950, and batch size to 16 for gradients accumulattion in Eq. \ref{eq:I}. Ideally, a larger batch size will give a better approximation of the importance score, but we find 16 is enough. When implement pruning, we select the minimum pruning proportion that effectively eliminates backdoors. In our experiments, Diff-Pruning removes 13\% of the channels per layer for DDPM and NCSN models trained on CIFAR-10, 7\% for DDPM models trained on CelebA and CelebA-HQ, and 5\% for LDM models trained on CelebA-HQ. In contrast, Diff-Cleanse prunes only 1\% to 2\% of the channels per layer across all models and samplers.

After pruning, the models are fine-tuned using synthesized data. We generate images through clean sampling across all models and randomly select images for fine-tuning: 6000 images for CIFAR-10 models, 6000 for CelebA, and 3000 for CelebA-HQ. All models are fine-tuned for 40 epochs with a learning rate of 2e-4. We set the batch size to 128 for DMs with $3\times 32\times 32$ latent space, 50 for $3\times 64\times 64$ space, and 16 for $3\times 256\times 256$ space because of GPU memory limitation.

\subsection{Hardware, Complexity and Time Cost}
The main experiments are conducted on a server equipped with Intel Golden 6240 CPUs and NVIDIA V100 GPUs.

In previous sections, we illustrate the outstanding performances of Diff-Cleanse in various settings. Here, we analyze the complexity of Diff-Cleanse to investigate its usefulness in practical applications, which is the sum of those in two stages. For stage 1, our proposed trigger inversion method BBTI, the time cost of calculating objective $\ell(\boldsymbol{g}^s)$ and MSCR score is $O(1)$. If we denote the number of iterations for the backdoor sampling as $m_1$, and the number of iterations for the optimization as $k$, the computational complexity for trigger inversion can be represented as $O(km_1)$. For stage 2, let the number of iterations for the gradients accumulation as $m_2$. The time cost of calculating Taylor importance score and pruning channels is $O(1)$. So the computational complexity for backdoor removal is $O(m_2)$. Suming the results of both stages, the overall computational complexity for our method except fine-tuning is $O(km_1+m_2)$. 

We record the time required to train a diffusion model from scratch and the time consumed by Diff-Cleanse on the CIFAR-10 dataset, as shown in Tab. \ref{tab:timecost}. Trigger inversion for BadDiff is relatively slow due to the use of DDPM sampler, which requires 1000 steps. Although VillanDiff involves more optimization steps, it consumes less time because it can employ the DDIM sampler for backdoor sampling, which only requires 50 steps. TrojDiff is compatible with both DDPM and DDIM samplers. Thus, we utilize DDIM for backdoor sampling to expedite trigger reconstruction. Backdoor removal typically takes approximately one minute, with subsequent fine-tuning serving as a balance between model performance and time complexity. While increasing the number of fine-tuning epochs and dataset size can enhance model performance, we account for the defender's limited computational resources, keeping fine-tuning time significantly shorter than training the diffusion model from scratch on the entire clean dataset.

\begin{table}[htbp]
\small
\centering
\setlength{\tabcolsep}{3.5pt}
\caption{The time cost of Diff-Cleanse for DDPM models trained on the CIFAR-10 dataset. The time is recorded based on our experiments on a single V100 GPU.}
\label{tab:timecost}
\begin{tabular}{cccccc}
\hline
Attack & Training & BBTI & Backdoor Removal & Fine-tuning \\
 \midrule BadDiff & 25.21h & 8.41min & 60.74s & 40.43min \\
 \midrule VillanDiff & 51.47h & 5.15min & 58.85s & 42.74min \\
 \midrule TrojDiff & 18.39h & 2.17min & 53.64s & 38.48min \\
\hline
\end{tabular}
\end{table}

\section{E More Results of Ablation Study}

\subsection{Effect of Gradients Accumulation Threshold}
\begin{figure}  
\centering
\includegraphics[width=1\linewidth]{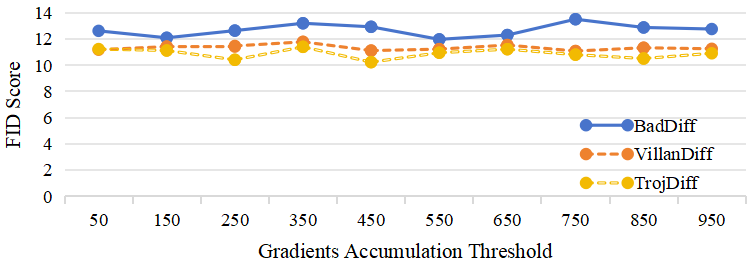}
\caption{The results of Diff-Cleanse using different threshold $\mathcal{T}$ in Eq. \ref{eq:I} against threee attacks. Models attacked by BadDiff and VillanDiff use “Stop Sign” trigger and “Shift” target. The model attacked by TrojDiff uses D2I attack, with blend-based trigger and “Mickey” target. Backdoors are completely removed from all models ($\text{ASR}=0$)}
\label{fig:gradientsaccumulationthresholdtrojdiff}
\end{figure}

Fig. \ref{fig:gradientsaccumulationthresholdtrojdiff} presents the results of Diff-Cleanse with various threshold $\mathcal{T}$ against BadDiff, VillanDiff and TrojDiff, demonstrating that our approach is not sensitive to $\mathcal{T}$. Our experiments across the three attack methods yield consistent results, indicating that channels sensitive to the trigger remain stable during sampling. Considering that the selection of $\mathcal{T}$ has little impact on the effectiveness of Diff-Cleanse, we use $\mathcal{T}=950$ in experiments outside the ablation study to accelerate computation.

\begin{figure}  
\centering
\includegraphics[width=0.8\linewidth]{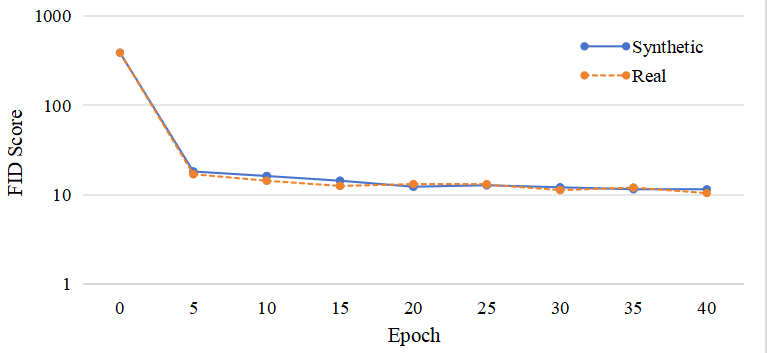}
\caption{The results of Diff-Cleanse using synthetic or real data on a model attack by VillanDiff, with “Stop Sign” trigger and “Shift” target.}
\label{fig:realdata}
\end{figure}

\begin{figure}  
\centering
\includegraphics[width=0.8\linewidth]{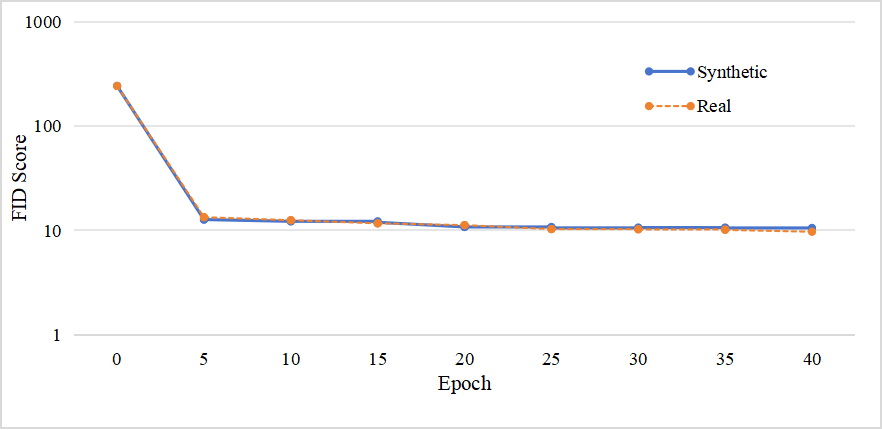}
\caption{The results of Diff-Cleanse using synthetic or real data on a model attack by TrojDiff, with blend-based trigger and “Mickey” target.}
\label{fig:realdata2}
\end{figure}

\subsection{Backdoor Removal with Real/Synthetic Data}
Fig. \ref{fig:realdata} and Fig. \ref{fig:realdata2} present the results of Diff-Cleanse using either synthetic or real data for fine-tuning, demonstrating that there is no significant difference in the training results between the two approaches. In both cases, the model's clean utility is effectively restored, indicating that our real-data-free backdoor removal approach is feasible. Since our trigger inversion and backdoor detection method is also sample-free, the entire framework can operate without requiring access to real data.

\section{F More Visual Examples}
\label{appendix:morevisualexamples}

\begin{figure}
\centering
\includegraphics[width=1\linewidth]{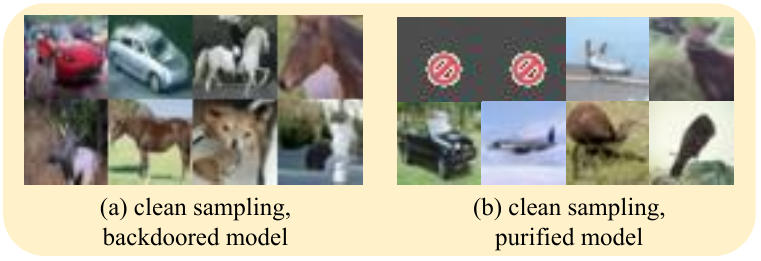}
\caption{Clean sampling results of a backdoored model and its puried version. The model is backdoored by BadDiff with trigger “Stop Sign” and “Shift”. For backdoor removal, we use a unstructural pruning method, which sets channels that are sensitive to the trigger to 0.}
\label{fig:unstructuralpruning}
\end{figure}

In this section, we provide additional visual examples. 

\paragraph{Visual Results of Unstructural Pruning.} Fig. \ref{fig:unstructuralpruning} illustrates the clean sampling results of a backdoored model and its corresponding purified model. We note that the backdoor in the model is not removed, but is altered. The target image even emerges during clean sampling. This observation push us to adopt structural pruning, which modifies the model's architecture and effectively mitigates the backdoor.

\begin{figure}
\centering
\includegraphics[width=0.8\linewidth]{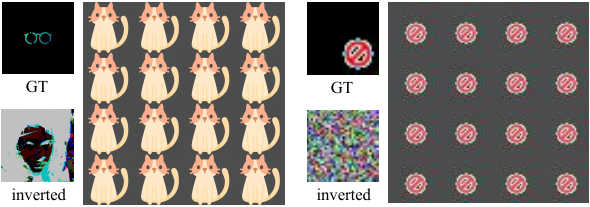}
\caption{Ground truth triggers “Glasses” and  “Stop Sign”, and their corresponding inverted triggers, as well as 16 generated images using inputs with the inverted triggers.}
\label{fig:invertedtrigger}
\end{figure}

\paragraph{Visual Results of Trigger Inversion.} Fig. \ref{fig:invertedtrigger} visualizes some ground truth triggers and their corresponding inverted triggers. In our threat model, defenders lack prior knowledge of attack methods. Our method therefore does not limit the color, size, and other visual properties of inverted triggers. Therefore, the inverted triggers are visually different from the ground truth of triggers. However, inverted triggers' ability to activate backdoors are strong, as shown in Tab. \ref{tab:triggerinversion}. The visual results of inverted triggers also demonstrate that when an attacker inserts a backdoor into the model, it creates a trigger space with multiple instances available.

\begin{figure}
\centering
\includegraphics[width=0.8\linewidth]{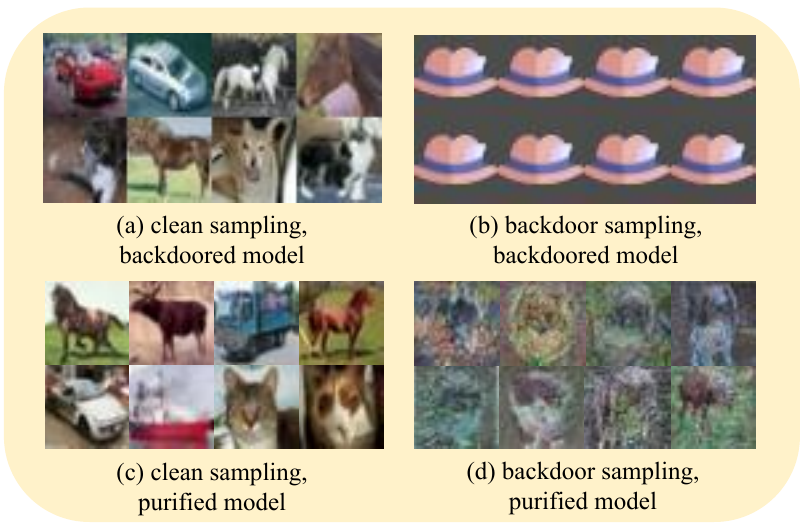}
\caption{The sample results of a backdoored DDPM model before (a-b) and after purification of Diff-Cleanse (c-d). The model is attacked by VillanDiff, trained on CIFAR-10 dataset with the trigger “Grey Box” and the target “Hat”. }
\label{fig:morevisualexamples1}
\end{figure}

\begin{figure}
\centering
\includegraphics[width=0.8\linewidth]{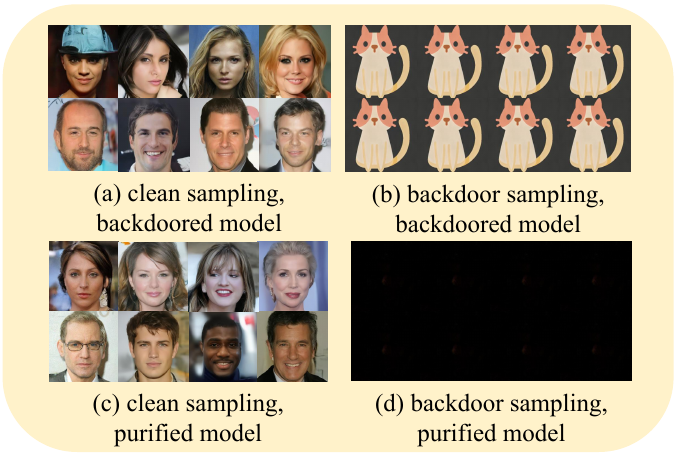}
\caption{The sample results of a backdoored DDPM model (a-b) and the purified diffusion model (c-d). The model is attacked by VillanDiff, trained on CelebA-HQ with the trigger “Glasses” and the target “Cat”. }
\label{fig:morevisualexamples2}
\end{figure}

\begin{figure}
\centering
\includegraphics[width=0.75\linewidth]{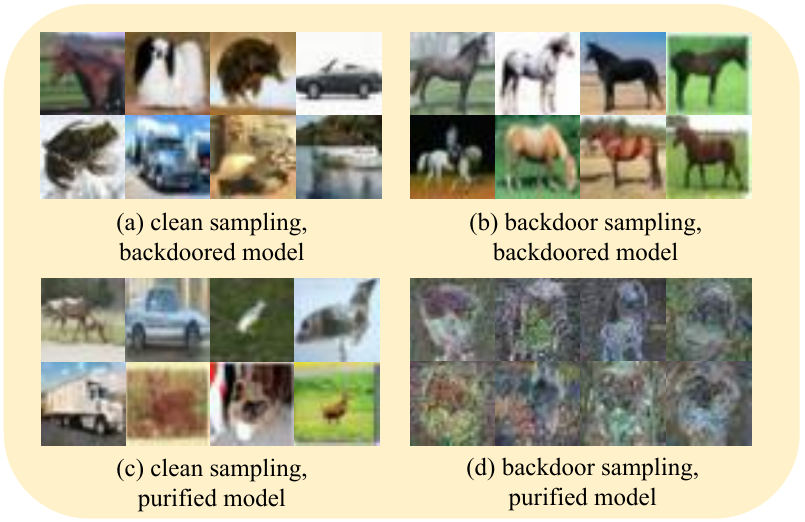}
\caption{The sample results of a backdoored DDPM model (a-b) and the purified diffusion model (c-d). The model is attacked by Din attack in TrojDiff, trained on CIFAR-10 with the blend-based trigger and the target “horse”. }
\label{fig:morevisualexamples3}
\end{figure}

\paragraph{Visual Results of Backdoor Removal.} Fig. \ref{fig:morevisualexamples1}, \ref{fig:morevisualexamples2} and \ref{fig:morevisualexamples3} illustrate the sample results of backdoored diffusion models before and after purified by Diff-Cleanse. After model repair, the model loses its ability to respond to the trigger and generate target images, while its clean utility remains unaffected.

\begin{figure}
\centering
\includegraphics[width=0.7\linewidth]{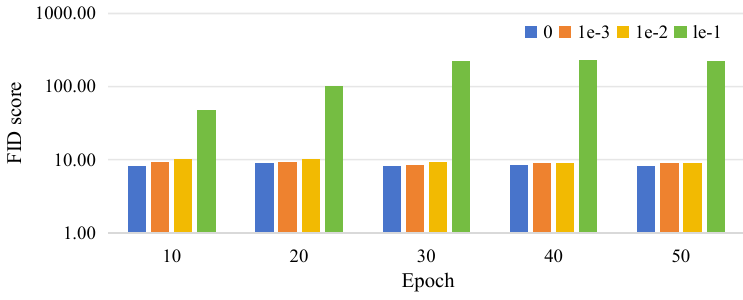}
\caption{FID scores during the training process of the adaptive attack, with the trigger “Grey Box”, the target “Hat”, and $\lambda=0,\ 1e-3,\ 1e-2, \ 1e-1$).}
\label{fig:adaptiveattacklambda}
\end{figure}

\section{G Details of Adaptive Attack}
\label{appendix:adaptiveattack}
After realizing the criteria of Diff-Cleanse, the attacker may train more concealed backdoor diffusion models, where the difference in activation values of clean noise and backdoored noise is reduced. To simulate this situation, we design an adaptive version of VillanDiff. Specifically, we calculate the Euclidean distance between the activation of clean noise and backdoored noise at the same time step, and use it as a weighted term in the loss. Such an adaptive attack aligns the activation in clean sampling and backdoor sampling, making the backdoor-related channels more stealthy.

As Eq. \ref{eq:adaptiveattack} shows, the proposed adaptive attack method adds a weighted item to the original loss $L_\theta$.
\begin{equation}
L^{\prime}_\theta(\boldsymbol{x}, t, \boldsymbol{\epsilon}, \boldsymbol{g})=L_\theta(\boldsymbol{x}, t, \boldsymbol{\epsilon}, \boldsymbol{g})+\lambda \left\| \Delta a_t \right\|^2
\label{eq:adaptiveattack}
\end{equation}
where $\Delta a_t$ is the difference between model outputs of the clean inputs and that of the backdoored inputs at time step $t$ and $\left\|\cdot\right\|$ is the $L-2$ norm. 

Then we attack diffusion models with the modified VillanDiff, which adopts adaptive attack loss $L^{\prime}_\theta$ and various $\lambda$. We train the diffusion models 50 epochs and evaluate them every 10 epochs. Fig. \ref{fig:adaptiveattacklambda} shows the utility of models during training process, indicating that the new loss term has limited effect on model's utility. However, when $\lambda=1e-1$, the model's FID score increases continuously during training, indicating that excessive alignment of the activations can lead to training failure. With $\lambda=1e-2,\ 1e-3$, the diffusion models lose little clean generation ability. Except for $\lambda=1e-1$, we implant backdoors into all other diffusion models and Tab. \ref{tab:adaptiveattack} shows the defence performance of Diff-Cleanse. Although the adaptive attack method makes the backdoor-related channels more concealed, our framework can still remove the backdoor and achieve good image quality in clean sampling.

\begin{table}[htbp]
  \centering
  \caption{The performance of Diff-Cleanse against the adaptive attack method. We use modified BadDiff and train the models on Cifar-10, with the trigger “Grey Box” and the target “Hat”.}
  \label{tab:adaptiveattack}
\begin{tabular}{cccc}
\hline
$\lambda$ & Detected & $\Delta \mathrm{FID}\downarrow$ & $\Delta \mathrm{ASR}\downarrow$ \\ \midrule
0 & $\checkmark$ & 2.24 & -1.00  \\
1e-3 & $\checkmark$ & 3.78 & -1.00 \\
1e-2 & $\checkmark$ & 2.95 & -1.00 \\
\hline
\end{tabular}
\end{table}

\end{document}